\documentclass[a4paper,10pt]{iopart}
\usepackage{hyperref}
\hypersetup{
pdfauthor = {Manuel Alberto Ballester S�chez},
pdftitle = {Estimating the spectrum of a density matrix with LOCC},
pdfsubject = {Quantum Estimation},
pdfkeywords = {spectrum estimation, entanglement estimation, depolarizing channel, Pauli channel},
pdfcreator = {LaTeX with hyperref package},
pdfproducer = {pdfLaTeX},
colorlinks={false},
pdffitwindow={true},plainpages={false}
}
\usepackage{amsfonts}
\usepackage{amssymb}
\usepackage{amsthm}

\usepackage{bm}
\usepackage{bbm}

\newcommand{\ran}{\rangle}
\newcommand{\lan}{\langle}
\newcommand{\be}{\begin{eqnarray}}
\newcommand{\ee}{\end{eqnarray} }
\newcommand{\benn}{\begin{eqnarray*}}
\newcommand{\eenn}{\end{eqnarray*} }
\newcommand{\mse}{\mathop{\mathrm{MSE}} \nolimits}
\newcommand{\esp}{\mathbb{E}}
\newcommand{\iu}{i}
\newcommand{\su}{\mathfrak{su}}
\newcommand{\re}{\mathop{\mathbb{R}}\nolimits}
\newcommand{\norm}[1]{\| #1\|}
\newcommand{\im}{\mathbbmss{1}}

\newcommand{\smalloh}{\mathrm{o}}
\newtheorem{theorem}{Theorem}

\newtheorem{lemma}[theorem]{Lemma}

\newcommand{\finproof}{$\Box$}
\newcommand{\bin}{\textrm{Bin}}

\newcommand{\rlprt}{\mathop{\mathrm{Re}} \nolimits}
\newcommand{\todst}{\stackrel{D}{\to}}
\newcommand{\var}{\mathop{\mathrm{Var}} \nolimits}
\newcommand{\hil}{\mathcal{H}}
\begin{document}
\title{Estimating the spectrum of a density matrix with LOCC}
\author{Manuel A. Ballester}
\address{Department of Mathematics, University of Utrecht, Box 80010, 3508 TA Utrecht, The Netherlands.\\
homepage: \href{http://www.math.uu.nl/people/balleste/}{http://www.math.uu.nl/people/balleste/}}
\ead{\mailto{ballester@math.uu.nl}}
\bibliographystyle{hapalike}
\begin{abstract}
The problem of estimating the spectrum of a density matrix is considered. Other problems, such as bipartite pure state entanglement, can be reduced to spectrum estimation. A local operations and classical communication (LOCC) measurement strategy is shown which is asymptotically optimal. This means that, for a very large number of copies, it becomes unnecessary to perform collective measurements which should be more difficult to implement in practice.
\end{abstract}

\section{Introduction}
Estimating a mixed state density matrix optimally, when one has $N$ copies of it available, is a difficult problem. The problem has been solved for qubits by \cite{VidalLatorre:optminmeas}, \cite{BBMT:collvslocal} and by \cite{HayashiMatsumoto:optestqbit} and it is known that optimal collective measurements perform strictly better than any measurement which can be implemented with local operations and classical communication (LOCC). For mixed qudits, i.e., mixed states on a Hilbert space of dimension $d$, not much work on finding optimal collective measurements has been done. In the present work a simpler case is studied, the estimation of the spectrum of a qudit density matrix. This problem has already been studied from the large deviation point of view by \cite{keylwerner:estspctr} and for the qubit case by \cite{Baganetal05a}.

In addition to being interesting in itself, spectrum estimation is useful because other problems can be reduced to it:
\begin{itemize}
\item Estimation of bipartite pure state entanglement. This problem has been studied for $d=2$ by \cite{SanchoHuelga00a} and by \cite{acin:entanglementestimation}.
\item Estimation of generalized Pauli channel. This problem has been studied by  \cite{Fujiwara:EstimationPauliChannel} and the depolarizing channel (special case of Pauli channel) by \cite{sasaki:estimationdepolchan}.
\end{itemize}

In the present paper, an LOCC asymptotically optimal\footnote{i.e. it performs asymptotically as well as any other measurement strategy} strategy will be described. The optimality of this LOCC strategy will be established by showing that it asymptotically satisfies the quantum Cram\'er-Rao bound (QCRB), stated by \cite{helstrom:book}. The QCRB is a bound on the mean square error of ``reasonable'' estimators. 

This paper is organized as follows. In section \ref{sec:prelis} the necessary concepts are introduced and it is specified what is meant by optimality. In section  \ref{sec:eststrat} the estimation strategy is described and the main result is stated more precisely (equation (\ref{eq:crbachiev})). In section \ref{sec:mseadaptive} the conditional mean square error matrix (MSE) is calculated, this is needed for the next two sections. A heuristic argument supporting the main result is given in section \ref{sec:heuristic} and a proof will be given in section \ref{sec:rigorous} (theorem \ref{theorem:mainresult}).

\section{Preliminaries}\label{sec:prelis}
The density matrix $\rho$ ($\rho\geq 0, \tr\rho =1$) will be parametrized in the following way:
$$
\rho(p)=\sum_{k=1}^{d-1} p_k |k\ran \lan k| + (1-\sum_{l=1}^{d-1} p_l) |d\ran \lan d|,
$$
where $p\in\Theta \subset \re^{d-1}$ is the parameter of interest, 
\benn
\Theta=\left\{ (p_1,\ldots,p_{d-1}):  0\leq p_k \leq 1, \sum_{k=1}^{d-1} p_k \leq 1 \right\}
\eenn
is the set of possible values of the parameter, and $\{|1\ran,\ldots,|d\ran\}$ is a basis of eigenvectors. 

The quantum estimation problem that will be studied in this paper is that, given $N$ copies of a completely unknown $\rho$, one is only interested in estimating its eigenvalues. Some of the needed concepts and results will be introduced next for the $N=1$ case.

Let $M$ be a measurement with outcomes in a finite set $\Omega$, i.e., a collection of matrices $\{M_\xi: \xi \in \Omega\}$ satisfying $M_\xi\geq 0$ and $\sum_{\xi \in \Omega}M_\xi=\im$, and let $\hat{p}=(\hat{p}_1,\ldots,\hat{p}_{d-1})$ be an estimator of $p$, i.e., a map from $\Omega$ to $\Theta$. The performance of such a measurement-estimator pair will be quantified by the MSE
$$
\mse(\hat{p},p,M)_{kl}=\esp[(\hat{p}_{k}-p_k)(\hat{p}_{l}-p_l)]=\sum_{\xi\in\Omega}\tr[\rho(p)M_\xi] (\hat{p}_{\xi k}-p_k)(\hat{p}_{\xi l}-p_l),
$$
where $\esp f$ means expectation of $f$.

The QCRB states that any unbiased\footnote{Unbiased means that $$\esp \hat{p}_k =\sum_{\xi\in\Omega }\tr[\rho(p)M_\xi]\hat{p}_{\xi k}=p_k.$$} measurement-estimator pair $(\hat{p},M)$ of $p$ satisfies
$$
\mse(\hat{p},p,M) \geq H(p)^{-1},
$$
where $H$ is the quantum Fisher information (QFI) (see for example \cite{helstrom:book} or \cite{holevo:book}). The QFI can be defined as the matrix with elements
$$
H(p)_{kl}=\rlprt \tr [\rho(p) \lambda_k(p) \lambda_l(p)],
$$
where $\{\lambda_1(p),\ldots,\lambda_{d-1}(p)\}$ are the symmetric logarithmic derivatives (SLD). The SLD are defined as selfadjoint solutions to the equation
\be \label{eq:slds}
\partial_{k}\rho(p) = \frac{\rho(p) \lambda_k(p)+\lambda_k(p) \rho(p)}{2},
\ee
where $\partial_k$ means partial derivative with respect to $p_k$.  

The SLD for the model studied in this paper are easy to calculate, indeed, writing (\ref{eq:slds}) on the basis of eigenvectors we get
$$
\lan i|[|k \ran\lan k| - |d\ran \lan d|]|j\ran = \frac{p_i+p_j}{2}\lan i|\lambda_k(p)|j\ran,
$$
or
$$
\lambda_k(p)=\frac{|k \ran\lan k|}{p_k} - \frac{|d\ran \lan d|}{p_d}.
$$

From the SLD one can then calculate the QFI to get:
\benn
H(p)_{k l}= \frac{\delta_{kl}}{p_k}+\frac{1}{p_d},~k,l\in\{1,\ldots,d-1\}
\eenn
where $p_d=1-\sum_{l=1}^{d-1} p_l$, the inverse of $H$ is
\be \label{eq:hinv}
H(p)^{-1}_{k l}= p_k \delta_{kl}-p_k p_l,~k,l\in\{1,\ldots,d-1\}.
\ee

When one has $N$ copies of $\rho$, i.e., the model is of the form $\rho(p)^{\otimes N}$ the QCRB becomes
$$
\mse(\hat{p},p,M)^{(N)} \geq \frac{H(p)^{-1}}{N},
$$
and this bound is valid for \emph{any} measurement $M$ (i.e. LOCC or not), as long as the measurement-estimator pair $(\hat{p},M)$ is unbiased.

The class of unbiased estimators, however, is too restrictive since in most practical situations one deals with biased ones. \cite{GillLevit:Bernouilli} used a multivariate extension of an inequality due to \cite{vantrees:book} to prove a more general bound. From their result and an inequality due to \cite{BraunsteinCaves:PRL}, it can be shown  that, under some regularity conditions, if $\sqrt{N}(\hat{p}-p)\todst Z(p)$ then 
\be \label{eq:asympcrb}
\var Z(p)\geq H(p)^{-1}, 
\ee
where ``$\todst$'' means convergence in distribution. This means that the variance of the limiting distribution of any regular estimator  satisfies the QCRB.

\section{Estimation strategy} \label{sec:eststrat}

Suppose now that one knows the basis of eigenvectors, and let us consider the measurement with elements $M_k = |k\ran \lan k|$. For this measurement the probability of outcome $k$ is
$$
\tr [\rho(p) M_k] = p_k.
$$
Now suppose this measurement is performed on $N$ copies of $\rho$, let $N_k$ be the number of times that outcome $k$ was observed, then $\{N_1,\ldots,N_{d-1} \}$ have a multinomial distribution, i.e.,
$$
\Pr(N_1=n_1,\ldots,N_{d-1}=n_{d-1})=\frac{ N!}{\prod_{k=1}^d n_k!}\prod_{k=1}^d p_k^{n_k},
$$
where $n_d=N-\sum_{k=1}^{d-1}n_k$.
The estimator
$$
\hat{p}_{k}=\frac{N_k}{N}
$$
is unbiased and a simple calculation shows that its MSE equals  the inverse of the QFI divided by $N$ which means that it saturates the QCRB and therefore it is optimal.

This would be the whole story, except for the fact that we have assumed that the eigenbasis of $\rho$ is known. If the eigenbasis is not known one can try to use a two-step adaptive strategy such as the one considered by \cite{gillmassar:pra}. The idea is to  make an initial rough estimate of $\rho$ on an asymptotically vanishing fraction of the copies, e.g., $N^\mu$ with $0<\mu<1$. Let $\sigma$ be that initial estimate of $\rho$ and $|\psi_k\ran$ be its (not necessarily unique) eigenbasis. On the rest of the copies ($N-N^\mu$) of $\rho$, the measurement with elements $M_k=|\psi_k\ran \lan \psi_k|$ is performed. 

In the rest of this paper, it will be shown that this method is asymptotically optimal, i.e., it asymptotically achieves the QCRB:
\be \label{eq:crbachiev}
\lim_{N\to \infty} N \mse(\hat{p},p,M)^{(N)} = H(p)^{-1},
\ee
provided $\mu$ is chosen strictly larger than $1/2$. 

\section{The MSE in the adaptive scheme}\label{sec:mseadaptive}

Let $N_i=N^\mu$ and $N_f=N-N^\mu$ be the sample sizes for the first and second stages respectively. In the second stage, the probability of outcome $k$, given the initial estimate $\sigma$, is
$$
q_k=\tr M_k \rho(p)=\lan \psi_k|\rho(p)|\psi_k\ran.
$$
These probabilities are  also a random variable. 

Next the MSE of the second stage (i.e. assuming fixed $q$'s) will be calculated. A condition for obtaining (\ref{eq:crbachiev}) will be derived from it.

Just as before, let $N_k$ be the number of times that outcome $k$ is observed and let us estimate $p_k$ as
$$
\hat{p}_{k}=\frac{N_k}{N_f}.
$$
The expectation of this estimator conditioned on $\sigma $is
$$
\esp [\hat{p}_k|\sigma]=q_k,
$$
so that in general it is a biased estimator. A simple calculation shows that the MSE conditioned on the first rough estimate of $\rho$ is
\be \label{eq:adaptivemse}
 \esp[(\hat{p}_{k}-p_k)(\hat{p}_{l}-p_l)|\sigma] = \frac{q_k \delta_{kl} - q_k q_l}{N_f} +(p_k-q_k)(p_l-q_l),
\ee
the second term is the square of the bias, the MSE itself is
\benn
\mse(\hat{p},p,M)^{(N)}= \esp [\esp[(\hat{p}_{k}-p_k)(\hat{p}_{l}-p_l)|\sigma]].
\eenn

Comparing (\ref{eq:hinv}) and (\ref{eq:adaptivemse}) and using the fact that $N/N_f \to 1$ as $N\to \infty$, it is easy to see that in order to get (\ref{eq:crbachiev}) it is sufficient that 
\be \label{eq:result}
\lim_{N\to \infty}\esp[N(q_k-p_k)(q_l-p_l)]= 0.
\ee
Indeed, if this is true, then it also holds that $\esp[q_k] \to p_k$ and $\esp[q_k q_l] \to p_k p_l$.

\section{Heuristic argument} \label{sec:heuristic}
Suppose for simplicity, that all eigenvalues of $\rho$ are different, then one expects that after the first estimate, the eigenbasis of $\rho$ and the eigenbasis of $\sigma$ are related by a unitary matrix which is very close to the identity, i.e.,
$$
|\psi_k\ran = U|k\ran,
$$
with 
$$
U=\exp\left( \iu \sum_{\alpha=1}^{d^2-1} \eta_\alpha T_\alpha \right)=\e^{i \eta \cdot T},
$$
where $\{T_1,\dots,T_{d^2-1}\}$ is a basis of $\su(d)$ satisfying $\tr T_\alpha T_\beta =\delta_{\alpha \beta}$, $\eta \in \re^{d^2-1}$ and $\norm{\eta}$ is small. One can then expand $U$ in Taylor series about $\eta=0$,
$$
U= \im + \iu\eta \cdot T - \frac{1}{2}\left(\eta \cdot T\right)^2  + \smalloh(\norm{\eta}^2).
$$
For any decent initial estimation strategy, $\eta$ is expected to go to $0$ as $N\to \infty$ at a rate of $N_i^{-1/2}=N^{-\mu/2}$.

The expression for $q_k$ is
$$
q_k=\sum_{l}p_l |\lan l|U|k\ran|^2,
$$
and
$$
|\lan l|U|k\ran|^2 = \delta_{kl}+\lan l|\eta\cdot T|k\ran \lan k|\eta\cdot T|l\ran - 
\delta_{kl} \lan k|(\eta\cdot T)^2|k\ran + \smalloh(\norm{\eta}^2),
$$
therefore
$$
q_k-p_k=\lan k| (\eta \cdot T) \rho (\eta \cdot T)|k\ran - p_k \lan k|(\eta \cdot T)^2|k\ran + \smalloh(\norm{\eta}^2).
$$
From the previous expression and the fact that  $\eta$ goes to zero at the rate $N^{-\mu/2}$ one can expect that 
$$
\esp (q_k-p_k)^2 = \frac{c}{N^{2\mu}}+ \smalloh(N^{-2\mu}),
$$
where $c$ is a constant possibly depending on $p$. From the previous equation it follows that
\be\label{eq:preveq}
\lim_{n\to\infty} N\esp (q_k-p_k)^2=0,
\ee
if and only if $\mu>1/2$.
Now, using (\ref{eq:preveq}) together with the Cauchy-Schwartz inequality
$$
\left(\esp[N(q_k-p_k)(q_l-p_l)]\right)^2\leq\esp[N(q_k-p_k)^2]~\esp[N(q_l-p_l)^2],
$$
(\ref{eq:result}) follows.  As pointed out before, the desired result (\ref{eq:crbachiev}) is a consequence of (\ref{eq:result}).

\section{Rigorous argument} \label{sec:rigorous}
\subsection{Some intermediate results}
If  $\rho = \im/d$, then any basis chosen for the second stage will give $(q_k-p_k) = 0$, so in what follows it is assumed that $\rho \neq \im/d$, i.e., $\rho$ has at least two different eigenvalues.

The following intermediate result will be needed. Basically it states that if $\rho$ and $\sigma$ are close to each other, then so will be their eigenvalues and eigenspaces.
\begin{lemma}\label{theorem:spisclose}
Let
\benn
\rho&=\sum_{a=1}^n p_a \Pi_a,\\
\sigma&=\sum_{k=1}^d s_k |\psi_k\ran \lan \psi_k|,
\eenn
where $p_a \neq p_b$ for $a\neq b$, $2\leq n \leq d$ is the number of different eigenvalues and $\Pi_a$ is a projector onto the eigenspace corresponding to eigenvalue $p_a$, and let $d_a = \tr \Pi_a$ be the degeneracy of $p_a$, also let 
$$
\Delta = \min_a \min_{b\neq a} |p_a-p_b|>0.
$$
If 
$$
d_{HS}(\rho,\sigma)=\sqrt{\tr(\rho-\sigma)^2}\leq \delta < \frac{\Delta}{1+\sqrt{d}},
$$
then
\begin{enumerate}
\item \label{item:point1} $\forall a,k$
$$
 |p_a -s_k| \sqrt{\lan \psi_k|\Pi_a|\psi_k\ran}  \leq \delta,
$$
i.e., either $p_a$ is close to $s_k$ or $|\psi_k\ran$ is almost orthogonal to the eigenspace corresponding to $p_a$.
\item \label{item:point2} $\forall a$ $\exists k$ such that $|p_a-s_k|\leq \delta$ and $\forall k$ $\exists a$ such that $|p_a-s_k|\leq \delta$, i.e., every eigenvalue of $\sigma$ is close to an eigenvalue of $\rho$ and vice versa. Let $M_a=\{k: |p_a-s_k|\leq \delta \}$ and $m_a=|M_a|>0$. Note that $M_a \cap M_b = \emptyset$ for $a\neq b$.
\item \label{item:point3}Let $a\neq b$, then if $k \in M_b$, then $|p_a-s_k|\geq \Delta - \delta$ and
$$
\sqrt{\lan \psi_k|\Pi_a|\psi_k\ran}\leq \frac{\delta}{\Delta -  \delta},
$$
i.e., if $s_k$ is  within a distance $\delta$ of $p_b\neq p_a$, then $|\psi_k\ran$ is almost orthogonal to the eigenspace corresponding to $p_a$.
\item \label{item:point4}$m_a=d_a$, i.e., for $\delta$ small enough, the number of eigenvalues of $\sigma$ within a distance $\delta$ from $p_a$ is equal to the degeneracy of $p_a$.
\item \label{item:point5}$\forall k \in M_a$,
$$
|p_a-\lan \psi_k|\rho|\psi_k\ran|\leq c(\rho) \delta^2,
$$
where 
$$
c(\rho)= \frac{4(d-1)}{\Delta}.
$$
\end{enumerate}
\end{lemma}
The proof of this lemma is given in \ref{sec:proofosspisclose}.

Now the way in which the first rough estimation is done will be specified. For this part it is convenient to represent $\rho$ and $\sigma$ in the following way
\benn
\rho 		&= \frac{\im}{d} + \theta \cdot T,\\
\sigma 	&= \frac{\im}{d} + \hat{\theta} \cdot T.
\eenn
The initial measurement strategy (which will be called \emph{plain tomography}) is to divide the initial number of copies $N_i$ in $d^2-1$ groups of size $N_0=N_i/(d^2-1)$, and in group $\alpha\in\{1,\ldots,d^2-1\}$ perform the measurement 
$$M^{(\alpha)}_{\pm} = \frac{\im \pm T_\alpha}{2}.$$
The probabilities are
$$p^{(\alpha)}_{\pm} = \frac{1 \pm \theta_\alpha}{2}.$$
Let $w_{\alpha +}$ be the number of times that outcome $+$ was obtained, it is binomially distributed $w_{\alpha +} \sim \bin(N_0,(1+\theta_\alpha)/2)$. The estimator for $\theta_\alpha$ is taken to be
$$
\hat{\theta}_\alpha = 2 \frac{w_{\alpha +}}{N_0} -1.
$$
The following result holds:
\begin{lemma} \label{theorem:convinprob}
If $\mu>1/2$ then $\forall \epsilon>0$ and $\forall h\geq 0$
\be \label{eq:convinprob}
\lim_{N\to \infty} \left(N^h\Pr\left[\sqrt{N}|q_k-p_k|\geq \epsilon \right]\right)=0.
\ee
\end{lemma}
The proof of this lemma is given in \ref{sec:proofofconvinprob}.
\subsection{Proof of the main result}

\begin{theorem}\label{theorem:mainresult}
If $\mu>1/2$ then (\ref{eq:crbachiev}) holds.
\end{theorem}
\begin{proof} Let $X^{(N)}_k=\sqrt{N}(q_k-p_k)$, clearly $(X^{(N)}_k)^2\leq N$. All that needs to be proven is that
$$
\lim_{N\to \infty}\esp [X^{(N)}_k X^{(N)}_l]=0.
$$
We have that
\be \label{eq:absandcs}
|\esp [X^{(N)}_k X^{(N)}_l]|\leq\esp [|X^{(N)}_k X^{(N)}_l|] \leq \sqrt{\esp[(X^{(N)}_k)^2]\esp[(X^{(N)}_l)^2]},
\ee
where in the second inequality the Cauchy-Schwartz inequality  has been used. Now choose any $\epsilon>0$, 
\benn
\esp[(X^{(N)}_k)^2]&= \sum_{x \geq 0} x \Pr[(X^{(N)}_k)^2=x]\\
						&= \sum_{0\leq x <\epsilon^2} x \Pr[(X^{(N)}_k)^2=x]+\sum_{ x>\epsilon^2} x \Pr[(X^{(N)}_k)^2=x]\\
						&\leq \epsilon^2  \Pr[(X^{(N)}_k)^2<\epsilon^2]+N \Pr[(X^{(N)}_k)^2\geq \epsilon^2]\\
						&\leq \epsilon^2  +N \Pr[|X^{(N)}_k|\geq \epsilon],
\eenn
using now (\ref{eq:convinprob}) one gets that $\forall \epsilon >0$,
$$
\lim_{N\to\infty}\esp[(X^{(N)}_k)^2]\leq \epsilon^2,
$$
which implies that it must be zero; this fact and  (\ref{eq:absandcs}) imply (\ref{eq:result}) and therefore the desired result (\ref{eq:crbachiev}).
\end{proof}
We have proven something about the limit of the MSE, but (\ref{eq:asympcrb}) is a bound to the variance of the limiting distribution. However, since the limit of the MSE cannot be smaller than the variance of the limiting distribution (which in this case can easily be proven to be Gaussian) it follows that our estimator achieves the bound (\ref{eq:asympcrb}).
\section{Estimation of bipartite pure state entanglement}
A bipartite entangled pure state $|\psi_{AB}\ran\in \hil_A\otimes \hil_B$ can be written as (Schmidt's decomposition)
$$
|\psi_{AB}\ran=\sum_{k=1}^d \sqrt{p_k}~|k\ran \otimes |e_k\ran,
$$
where $\{|k\ran\}$ and $\{|e_k\ran\}$ are orthonormal basis of $\hil_A$ and $\hil_B$ which are both of dimension $d$.

The entanglement of $|\psi_{AB}\ran$ can be calculated as the entropy of one of the reduced states, 
$$
E(|\psi_{AB}\ran)=-\tr (\rho_A \log_2 \rho_A)=-\sum_{k=1}^d p_k \log_2 p_k,
$$
where $\rho_A = \tr_B |\psi_{AB}\ran\lan \psi_{AB}|$, i.e., entanglement is a function of the eigenvalues of the reduced density matrix. This means that entanglement can be estimated by performing measurements on $\rho_A$ only, in order to estimate its spectrum. The question is whether this procedure is optimal. A quick calculation of  the QFI for the parameters $p_k$ in the model given by $|\psi_{AB}\ran$ shows that indeed the entanglement of $|\psi_{AB}\ran$ can be optimally estimated by estimating the spectrum of $\rho_A$ using the procedure described above in this paper.

The same result\footnote{That entanglement can be optimally estimated by estimating the spectrum of the reduced density matrix.} was obtained by  \cite{acin:entanglementestimation} for $d=2$ using other tools.

\section{Conclusions}
The estimation of the spectrum of a finite dimensional density matrix has been analyzed. The following LOCC procedure has been studied:
\begin{enumerate}
\item Perform the so called plain tomography on $N^\mu$ copies where $\mu>1/2$ and $N$ is the total number of copies. From this one gets an initial estimate of the whole density matrix, call it $\sigma$. Let $|\psi_1\ran,\ldots,|\psi_d\ran$ be a set of eigenvectors of $\sigma$.
\item Perform the measurement with elements $M_k=|\psi_k\ran\lan \psi_k|$ on the remaining $N-N^\mu$ copies and estimate $p_k$ as the number of times the outcome $k$ was obtained divided by $N$.
\end{enumerate}
It has been shown that the above procedure performs asymptotically as well as any measurement (including collective ones). This means that in the asymptotic regime there is no need to perform the more complicated collective measurements for the estimation of the spectrum of a density matrix (or pure bipartite entanglement).

\ack
I would like to thank Richard Gill, Madalin Gu\c{t}\u{a} and Igor Grubi\u{s}i\'{c }for their very useful comments.
This research was funded by the Netherlands Organization for
Scientific Research (NWO), support from the RESQ (IST-2001-37559)
project of the IST-FET programme of the European Union is also
acknowledged.

\appendix
\section{Proof of lemma \ref{theorem:spisclose}}\label{sec:proofosspisclose}
\begin{enumerate}
\item The square of the distance between $\rho$ and $\sigma$ can be written as
\benn
d_{HS}(\rho,\sigma)^2&= \sum_{k=1}^d \sum_{a=1}^n \lan \psi_k|(\rho -\sigma)\Pi_a|(\rho -\sigma)|\psi_k\ran \\
 & = \sum_{k=1}^d \sum_{a=1}^n (p_a -s_k)^2 \lan \psi_k|\Pi_a|\psi_k\ran  \leq \delta^2.
\eenn
Since all terms are nonnegative, this implies that all of them are less than or equal to $\delta$ and this implies point  (\ref{item:point1}).

\item For point (\ref{item:point2}), only the first statement will be proven, the proof of the second is almost identical. Suppose that the opposite is true, i.e., that $\exists a$ such that $\forall k$ $|p_a-s_k|>\delta$ then
\benn
d_{HS}(\rho,\sigma)^2&= \sum_{k=1}^d \sum_{b=1}^n (p_b -s_k)^2 \lan \psi_k|\Pi_b|\psi_k\ran \\
										 &\geq \sum_{k=1}^d  (p_a -s_k)^2 \lan \psi_k|\Pi_a|\psi_k\ran \\
										 &> \delta^2 \tr \Pi_a \geq \delta^2,
\eenn
i.e., $d_{HS}(\rho,\sigma)>\delta$ which is a contradiction. 

\item $|p_a-s_k|=|(p_a-p_b)+(p_b-s_k)|\geq |p_a-p_b|-|p_b-s_k|\geq \Delta -\delta$, the second statement follows from the previous inequality and point (\ref{item:point1}).
\item 
\benn
m_a &= \sum_{k\in M_a} \lan \psi_k|\psi_k\ran \geq \sum_{k\in M_a} \lan \psi_k|\Pi_a|\psi_k\ran \\
		&=\tr \Pi_a -\sum_{k\notin M_a} \lan \psi_k|\Pi_a|\psi_k\ran\\
		&\geq \tr \Pi_a - \sum_{k\notin M_a} \left(\frac{\delta}{\Delta -\delta}\right)^2\\
		&\geq \tr \Pi_a - d \left(\frac{\delta}{\Delta -\delta}\right)^2,
\eenn
where point (\ref{item:point3}) has been used. Now, since $d_a=\tr \Pi_a$, we get
$$
m_a \geq d_a - d \left(\frac{\delta}{\Delta -\delta}\right)^2.
$$
Since $\delta < \Delta/(1+\sqrt{d})$, 
$$
d \left(\frac{\delta}{\Delta -\delta}\right)^2<1,
$$
and since $m_a$ is an integer, we have that $m_a\geq d_a$. Using the fact that $\sum_a m_a = \sum_a d_a = d$, we get that $m_a=d_a$.
\item Let $a\neq b$, and $k\in M_a$
\benn\fl
|p_a-p_b|\sqrt{\lan \psi_k|\Pi_b|\psi_k\ran}& =|(p_a-s_k)+(s_k-p_b)|\sqrt{\lan \psi_k|\Pi_b|\psi_k\ran}\\
																						&\leq [|p_a-s_k|+|s_k-p_b|]\sqrt{\lan \psi_k|\Pi_b|\psi_k\ran}\\
						&\leq \left[\delta\sqrt{\lan \psi_k|\Pi_b|\psi_k\ran}+|s_k-p_b|\sqrt{\lan \psi_k|\Pi_b|\psi_k\ran}\right]\\
						&\leq \left[\delta\sqrt{\lan \psi_k|\Pi_b|\psi_k\ran}+\delta\right]\leq 2\delta,
\eenn
where points (\ref{item:point1}) and (\ref{item:point2}) have been used.
Thus, we have that
\benn
\lan \psi_k|\Pi_b|\psi_k\ran \leq \frac{4 \delta^2}{(p_a-p_b)^2}.
\eenn

Now I turn to the quantity of interest, 
\benn
|p_a-\lan \psi_k|\rho|\psi_k\ran|&= \left | p_a- \sum_b p_b \lan \psi_k|\Pi_b|\psi_k\ran \right|\\
																&= \left |\sum_b (p_a-p_b) \lan \psi_k|\Pi_b|\psi_k\ran  \right|\\
																&\leq \sum_b |p_a-p_b| \lan \psi_k|\Pi_b|\psi_k\ran  \\
																&= \sum_{b\neq a} |p_a-p_b| \lan \psi_k|\Pi_b|\psi_k\ran \\
																& \leq 4  \sum_{b\neq a} \frac{1}{|p_a-p_b|}\delta^2\\
																&\leq \frac{4(d-1)}{\Delta} \delta^2=c(\rho) \delta^2.~\textrm{\finproof}
\eenn
\end{enumerate}
\section{Proof of lemma \ref{theorem:convinprob}} \label{sec:proofofconvinprob}
Now we enumerate the eigenvalues of $\rho$ from $1$ to $d$ again, with some of them possibly equal. Points (\ref{item:point2}) and (\ref{item:point4})  of lemma \ref{theorem:spisclose}, take care that for every eigenvalue of $\rho$, the right number of eigenvalues of $\sigma$ will satisfy point (\ref{item:point5}). From point (\ref{item:point5}) of lemma \ref{theorem:spisclose} we get that $|q_k-p_k|\geq c(\rho) \delta^2$ implies $d(\rho,\sigma)^2 \geq \delta^2$, we have
\benn
\Pr\left[|q_k-p_k|\geq c(\rho) \delta^2 \right]
		& \leq \Pr\left[d(\rho,\sigma)^2\geq \delta^2 \right]\\
		& = \Pr\left[\sum_{\alpha=1}^{d^2-1}(\theta_\alpha - \hat{\theta}_\alpha)^2 \geq \delta^2 \right].
\eenn
Since 
$$
\sum_{\alpha=1}^{d^2-1}(\theta_\alpha - \hat{\theta}_\alpha)^2 \geq \delta^2
$$
 implies that for at least one $\alpha$ 
$$(\theta_\alpha - \hat{\theta}_\alpha)^2 \geq \frac{\delta^2}{d^2-1},$$
it follows that 
\benn\fl
\Pr\left[\sum_{\alpha=1}^{d^2-1}(\theta_\alpha - \hat{\theta}_\alpha)^2 \geq\delta^2 \right]
  &\leq 1 - \Pr\left[\forall \alpha, (\theta_\alpha - \hat{\theta}_\alpha)^2 <  \frac{\delta^2}{d^2-1} \right]\\
  &=1-\prod_{\alpha=1}^{d^2-1}\Pr\left[ |\theta_\alpha - \hat{\theta}_\alpha| <  \frac{\delta}{\sqrt{d^2-1}} \right]\\
&=1-\prod_{\alpha=1}^{d^2-1}\Pr\left[ \left|w_{\alpha +} - \frac{1+\theta_\alpha}{2}N_0\right| < \frac{N_0}{2}\frac{\delta}{\sqrt{d^2-1}} \right]\\
&=1-\prod_{\alpha=1}^{d^2-1}\left(1-\Pr\left[ \left|w_{\alpha +} - \frac{1+\theta_\alpha}{2}N_0\right| \geq \frac{N_0}{2}\frac{\delta}{\sqrt{d^2-1}} \right]\right)\\
&\leq 1-\left(1-2 \exp\left[- \frac{\delta^2}{2(d^2-1)}N_0 \right]\right)^{d^2-1}.
\eenn
In the last inequality we have used a form of the Chernoff bound\footnote{If $X\sim\bin(n,p)$ then $\Pr[|X-np|\geq\lambda]\leq 2 \exp(-2 \lambda^2/n).$}. Thus, we finally have that
$$
\Pr\left[|q_k-p_k|\geq c(\rho) \delta^2 \right] \leq 1-\left(1-2 \exp\left[- \frac{\delta^2}{2(d^2-1)}N_0 \right]\right)^{d^2-1},
$$
now let $c(\rho) \delta^2=\epsilon N^{-1/2}$ and substitute $N_0$ by its value, $N^\mu/(d^2-1)$, the result is
\benn 
\Pr\left[\sqrt{N}|q_k-p_k|\geq \epsilon \right]&\leq 1-\left(1-2 \exp\left[-\frac{\epsilon N^{\mu-1/2}}{2 c(\rho)(d^2-1)^2}  \right]\right)^{d^2-1}\\
&\leq 2(d^2-1)\exp\left[-\frac{\epsilon N^{\mu-1/2}}{2 c(\rho)(d^2-1)^2}  \right],
\eenn
multiplying by $N^h$, taking $\mu>1/2$ and $N\to \infty$, we get the desired result (\ref{eq:convinprob}).

\end{document}